\documentclass[10pt,aps,prl,twocolumn,superscriptaddress,longbibliography]{revtex4-2}
\usepackage{placeins}

\usepackage{comment}
\usepackage[normalem]{ulem}
\usepackage[english]{babel}
\usepackage{graphicx}
\usepackage{dcolumn}
\usepackage{bm}
\usepackage{blindtext}
\usepackage{verbatim}
\usepackage{mathrsfs}
\usepackage{musicography}
\usepackage{amsmath}
\usepackage{dsfont}
\usepackage{mathrsfs}
\usepackage{amssymb}
\usepackage{amsthm}
\usepackage{cancel}
\usepackage{physics}
\usepackage{epstopdf}
\usepackage{mathtools}
\usepackage{color}
\usepackage{epsfig}
\usepackage{pst-grad} 
\usepackage{pst-plot} 
\usepackage{hyperref}
\usepackage{verbatim}
\usepackage{afterpage}
\usepackage{sidecap}
\usepackage{booktabs}
\usepackage{tabularx}
\usepackage{array}
\usepackage{graphicx}

\usepackage{adjustbox}
\hypersetup{
    colorlinks=true,
    linkcolor=blue,
    filecolor=red,      
    urlcolor=cyan,
}
\usepackage{tikzsymbols}

\begin{document}
\title{Is Exact Markovianity Fundamental Once Time Is Relational?}
\author{Partha Nandi}
\email{pnandi@sun.ac.za}
\affiliation{National Institute of Theoretical and Computational Sciences (NITheCS), Stellenbosch, 7604, South Africa}
\affiliation{Department of Physics, University of Stellenbosch, Stellenbosch, 7600, South Africa}

\author{Francesco Petruccione}
\email{petruccione@sun.ac.za}
\affiliation{National Institute of Theoretical and Computational Sciences (NITheCS), Stellenbosch, 7604, South Africa}
\affiliation{School of Data Sciewnce and Computational Thinking, Stellenbosch University, Stellenbosch 7600, South Africa}
\affiliation{Department of Physics, University of Stellenbosch, Stellenbosch, 7600, South Africa}

\begin{abstract}

Markovian open quantum theory assumes evolution with respect to an external classical time parameter, yet no preferred notion of time exists fundamentally in relativistic physics. We resolve this tension by formulating relativistic open quantum dynamics relationally through finite-resolution quantum clocks. Using the Schwinger--Tomonaga formalism, we derive a covariant master equation directly on spacetime hypersurfaces and show that the resulting reduced dynamics is generically non-Markovian even for local interactions. Environmental correlations and clock fluctuations jointly generate the memory kernel, while the Gorini--Kossakowski--Lindblad--Sudarshan (GKLS) structure emerges only after relational coarse-graining, implying that exact Markovian evolution is effective rather than fundamental. In the sharp-clock limit, the formalism reduces to the Anastopoulos--Hu gravitational decoherence equation.

\end{abstract}

\maketitle

\paragraph{Introduction.—}
One of the less-understood conceptual questions at the interface of quantum theory and relativity concerns the status of exact Markovian evolution when time is treated as a physical rather than an external concept. The Gorini--Kossakowski--Lindblad--Sudarshan (GKLS) framework assumes evolution with respect to an external classical time parameter and describes reduced dynamics through exact time-local Markovian semigroups~\cite{Gorini1976,Lindblad1976,Gorini:1976cm,PhysRevA.103.062226,Breuer:2015zlm}. Relativistic physics, however, admits no preferred global notion of time. Instead, microscopic evolution is formulated covariantly on spacelike hypersurfaces through the Schwinger-Tomonaga equation~\cite{Tomonaga1946,Schwinger1948}. Whether exact Markovian open-system dynamics can therefore be fundamentally compatible with a relational notion of time remains an unresolved question~\cite{BreuerPetruccione2000Relativistic,Hu:2020luk}.

The origin of this tension lies in the operational status of time itself. Conventional open-system dynamics assumes evolution with respect to an external classical time parameter~\cite{Jana:2021xyn}, whereas relativistic and gravitational physics admit only relational notions of temporal ordering defined through physical degrees of freedom. More generally, recent work has shown that quantum dynamics can be formulated in a manifestly Lorentz-covariant manner without introducing a preferred external notion of time~\cite{Nandi:2023tfq}. Relational approaches to quantum theory therefore argue that physical evolution should be formulated through correlations among physical degrees of freedom rather than with respect to an external temporal background~\cite{Giulini1996}. In the Page-Wootters construction~\cite{PhysRevD.27.2885}, temporal evolution emerges through conditioning on the state of a quantum clock, while Rovelli's relational framework~\cite{PhysRevD.43.442} emphasizes that physical predictions are fundamentally expressed as correlations among observables rather than evolution with respect to an external time parameter. Realistic-clock formulations further suggest that finite clock resolution can induce departures from exact unitary evolution even in otherwise closed quantum systems~\cite{PhysRevLett.93.240401}. Taken together, these developments indicate that physical time should be regarded as an operational concept arising from quantum correlations rather than a fundamental background structure.

However, while relational approaches explain how temporal evolution can emerge from physical clocks and realistic-clock models reveal consequences of finite clock accuracy for quantum coherence, the implications of finite clock resolution for the structure of relativistic open-system dynamics remain largely unexplored. In particular, if time is operationally defined through physical clocks rather than an exact external parameter, it is not obvious that the exact Markovian evolution underlying GKLS theory should remain fundamental.


Several observations suggest that the problem is not merely technical. Exact Lorentz-invariant Markovian semigroups are generically incompatible with locality and vacuum stability~\cite{Breuer1998RelativisticQSD,Diosi2022,Breuer1998ReplyQSD}, while relativistic causality imposes strong constraints on covariant extensions of GKLS dynamics~\cite{Haag:1992hx}. These results suggest that relativistic non-Markovianity may not simply reflect environmental complexity or approximation schemes, but may instead originate from the operational structure of relativistic time itself.

Recent works have also investigated relativistic open-system dynamics and covariant GKLS structures in scattering and quantum-field-theoretic settings~\cite{PhysRevD.111.045014}. Such approaches demonstrate that effective relativistic GKLS dynamics can emerge in asymptotic scattering regimes. The present work differs conceptually in that non-Markovianity arises operationally from relational clock conditioning itself, rather than from subsystem reduction with respect to an external classical time parameter.

\begin{figure*}[!t]
\centering
\includegraphics[width=\textwidth]{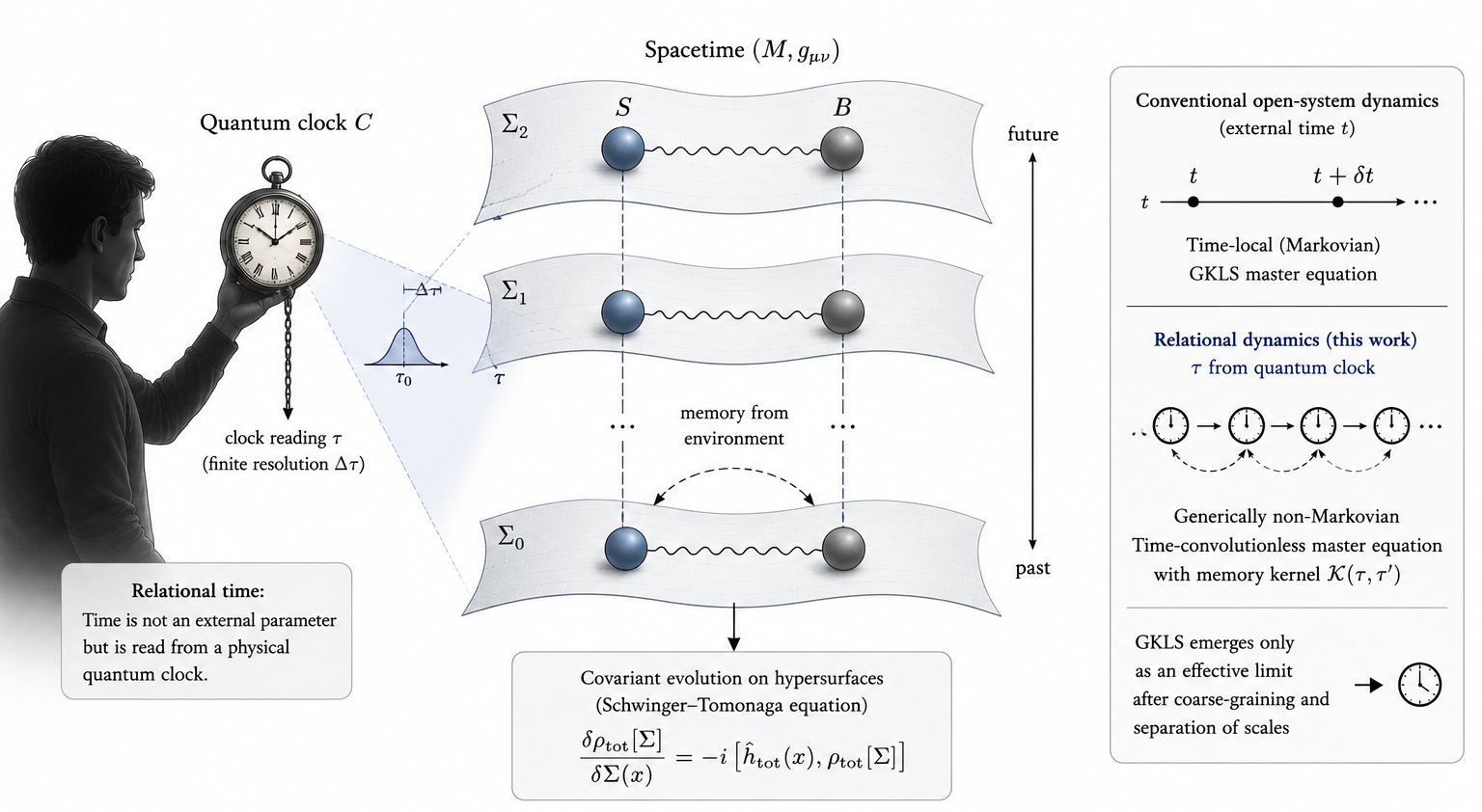}
\vspace{-2mm}
\vspace{-3mm}
\caption{
Covariant relational open-system dynamics.
A physical observer defines operational time by reading a finite-resolution quantum clock rather than referring to an external classical time parameter. The standing observer on the left represents an operational agent who extracts relational time~$\tau$ from the quantum clock, whose finite temporal resolution $\Delta\tau$ introduces intrinsic temporal uncertainty. The subsystem $S$ interacts locally with its environment $B$ on spacetime hypersurfaces $\Sigma$, while the total state evolves covariantly according to the Schwinger--Tomonaga equation. Conditioning the dynamics on the clock reading induces a relational reduced evolution in which memory effects arise jointly from environmental correlations and clock fluctuations. In contrast to conventional GKLS dynamics defined with respect to an external coordinate time~$t$, the relational dynamics is generically non-Markovian even for local interactions. Conventional Markovian evolution emerges only after relational coarse-graining and in the sharp-clock limit where clock fluctuations become negligible.
Operational time extracted from a realistic finite-resolution quantum clock generically makes reduced quantum dynamics non-Markovian, implying that exact Markovianity is an emergent effective limit rather than a fundamental property of relativistic open quantum systems.
}
\label{fig:relational}
\vspace{-3mm}
\end{figure*}

In this letter we resolve this problem by developing a covariant relational framework for open quantum dynamics in which evolution is conditioned on finite-resolution quantum clocks. Starting from a closed subsystem--environment--clock model governed by Schwinger--Tomonaga evolution, we derive a covariant time-convolutionless master equation directly on spacetime hypersurfaces. In the present framework, the decomposition into subsystem, environment, and clock sectors is not assumed to be fundamentally unique, but reflects an operational separation between the degrees of freedom being probed, inaccessible environmental correlations, and the physical subsystem used to define temporal ordering. A relational clock is therefore understood as a physical degree of freedom whose correlations provide an approximately stable timelike ordering for conditioned subsystem evolution. Different relational clock choices may in principle lead to different effective reduced descriptions, although physically reasonable clock sectors are expected to generate approximately consistent coarse-grained dynamics whenever clock fluctuations remain sufficiently small. A schematic representation of the relational framework is shown in Fig.~\ref{fig:relational}.

Unlike recent proposals involving fundamentally nonlinear modifications of Tomonaga--Schwinger dynamics~\cite{Hsu2026,Diosi2026}, the present framework preserves linear covariant microscopic evolution, with non-Markovianity emerging only after relational conditioning and subsystem reduction.

Our central result is that finite clock resolution generically induces non-Markovian reduced dynamics even for local subsystem--environment interactions. Environmental correlations and relational clock fluctuations jointly generate the memory kernel, implying that temporal nonlocality originates not only from environmental structure but also from the operational definition of time itself. We further show that the familiar GKLS structure emerges only after relational coarse-graining together with a separation of scales between bath memory and clock resolution, so that exact Markovian evolution appears as an effective rather than fundamental feature of relativistic open-system dynamics.

Finally, as a consistency check, we recover the Anastopoulos--Hu gravitational decoherence equation~\cite{AnastopoulosHu2013} in the sharp-clock fixed-foliation limit. Conventional coordinate-time master equations therefore emerge as limiting cases of a more general relationally non-Markovian dynamics in which relational clock fluctuations become negligible and time effectively behaves classically.

\paragraph{Relational covariant dynamics.—}
In a generally covariant setting, no preferred external time parameter exists fundamentally. Physical evolution must therefore be defined relationally through correlations among dynamical degrees of freedom rather than with respect to an underlying background notion of time. In such a framework, clock variables are not external classical parameters but physical quantum degrees of freedom whose correlations operationally parameterize subsystem evolution. Following relational approaches based on evolving observables~\cite{PhysRevD.43.442}, physical observables are defined conditionally relative to clock readings rather than with respect to an external coordinate time. Relational dynamics is therefore formulated directly in terms of correlations among observables on spacetime hypersurfaces.

The decomposition into subsystem, environment, and clock sectors is not fundamental, but reflects an operational separation between the degrees of freedom being probed, inaccessible environmental degrees of freedom, and the physical subsystem used as a temporal reference. A degree of freedom becomes identifiable as a relational clock when its correlations define an approximately stable timelike ordering structure for conditioned subsystem evolution.

We therefore consider a composite relativistic quantum system on a globally hyperbolic spacetime $(M,g_{\mu\nu})$, where $M$ denotes the spacetime manifold and $g_{\mu\nu}$ the spacetime metric. The total system consists of a subsystem sector $S$, an environment (bath) sector $B$, and a relational quantum clock sector $C$. Here $S$ denotes the physical subsystem whose reduced dynamics is studied, $B$ represents inaccessible environmental degrees of freedom, and $C$ is a physical quantum clock whose readings operationally define relational time.

In a relativistic setting, interactions must be formulated locally on spacetime in order to preserve covariance and causal consistency. We therefore describe the coupled dynamics through local operator densities defined on spacelike hypersurfaces $\Sigma$. The subsystem and environment may in general represent relativistic quantum fields or localized degrees of freedom interacting through local spacetime couplings. The total Hilbert space is
\begin{equation}
\mathcal H_{\rm tot}
=
\mathcal H_S
\otimes
\mathcal H_B
\otimes
\mathcal H_C.
\end{equation}

The total state evolves covariantly across spacelike hypersurfaces according to the Schwinger--Tomonaga equation~\cite{Tomonaga1946,Schwinger1948},
\begin{equation}
\frac{\delta \rho_{\rm tot}[\Sigma]}
{\delta\Sigma(x)}
=
-i
[
\hat h_{\rm tot}(x),
\rho_{\rm tot}[\Sigma]
],
\label{eq:ST}
\end{equation}
where $\Sigma$ denotes a spacelike hypersurface and $\hat h_{\rm tot}(x)$ is the local Hamiltonian density,
\begin{equation}
\hat h_{\rm tot}(x)
=
\hat h_S(x)
+
\hat h_B(x)
+
\hat h_C(x)
+
g
\sum_\alpha
\hat A_\alpha(x)
\otimes
\hat B_\alpha(x).
\label{eq:htot}
\end{equation}
Here $g$ denotes a weak subsystem--environment coupling constant, while $\hat A_\alpha(x)$ and $\hat B_\alpha(x)$ are local spacetime operators associated respectively with the subsystem and environmental sectors, with $\alpha$ labeling different interaction channels. The microscopic Schwinger--Tomonaga evolution therefore remains local and covariant, while temporal nonlocality emerges only at the level of conditioned reduced dynamics after relational conditioning and reduction over inaccessible degrees of freedom.

Microcausality,
\begin{equation}
[
\hat h_{\rm tot}(x),
\hat h_{\rm tot}(y)
]
=
0,
\qquad
(x-y)^2<0,
\end{equation}
ensures integrability of the hypersurface evolution.

Relational time is introduced operationally through conditioning on finite-resolution quantum clock readings. Rather than evolving with respect to an external classical time parameter, the reduced dynamics is defined relative to correlations with a physical clock degree of freedom. The relevant notion of time therefore emerges from conditional relations among observables on spacetime hypersurfaces. In the present framework, a relational clock is characterized by the existence of sufficiently coherent clock correlations capable of defining an approximate causal ordering structure for subsystem evolution. Finite clock resolution then implies that this ordering is operationally smeared over neighboring relational clock readings.

To implement this operationally, we introduce a quantum clock instrument
\begin{equation}
\mathcal J^{(\sigma)}_{\Sigma,\tau}:
\mathcal T_1(\mathcal H_{\rm tot})
\rightarrow
\mathcal T_1(\mathcal H_{\rm tot}),
\end{equation}
which conditions the total state on the relational clock reading $\tau$ at the hypersurface $\Sigma$. Here $\mathcal T_1(\mathcal H_{\rm tot})$ denotes the trace-class operators on the total Hilbert space. The map is completely positive and trace nonincreasing, reflecting the probabilistic nature of relational conditioning.

The corresponding conditioned reduced state is
\begin{equation}
\widetilde\rho_{S,\tau}[\Sigma]
=
\Tr_{B,C}
\Big[
\mathcal J^{(\sigma)}_{\Sigma,\tau}
(
\rho_{\rm tot}[\Sigma]
)
\Big],
\label{eq:conditionedstate}
\end{equation}
where the trace is taken over the bath and clock sectors. The normalized conditioned state is then
\begin{equation}
\rho_{S,\tau}[\Sigma]
=
\frac{
\widetilde\rho_{S,\tau}[\Sigma]
}{
\Tr
\widetilde\rho_{S,\tau}[\Sigma]
}.
\end{equation}

Complete positivity therefore applies at the level of the conditioned map and subnormalized state, while the normalized conditioned state evolves nonlinearly due to relational conditioning.

Passing to the interaction picture with respect to the uncoupled subsystem, bath, and clock sectors yields
\begin{equation}
\frac{\delta \rho^I_{\rm tot}[\Sigma]}
{\delta\Sigma(x)}
=
-i
[
\hat h_I^I(x),
\rho^I_{\rm tot}[\Sigma]
],
\end{equation}
with interaction Hamiltonian density
\begin{equation}
\hat h_I^I(x)
=
g
\sum_\alpha
\hat A^I_\alpha(x)
\otimes
\hat B^I_\alpha(x).
\end{equation}

Iterating the Schwinger--Tomonaga equation perturbatively generates a covariant Dyson expansion over causally ordered spacetime regions. Applying the clock instrument and tracing over the bath and clock sectors yields a covariant time-convolutionless master equation for the conditioned reduced state (see Supplemental Material),
\begin{align}
\frac{\delta \widetilde\rho^I_{S,\tau}[\Sigma]}
{\delta\Sigma(x)}
={}&
-g^2
\sum_{\alpha\beta}
\int_{J^-(x)\cap\Omega(\Sigma_0,\Sigma)}
\dd\mathrm{vol}_y
\Big(
\mathcal C^{(\sigma)}_{\alpha\beta}(x,y;\tau)
\nonumber\\
&\times
[
\hat A^I_\alpha(x),
\hat A^I_\beta(y)
\widetilde\rho^I_{S,\tau}[\Sigma]
]
\nonumber\\
&+
\mathcal C^{(\sigma)*}_{\beta\alpha}(y,x;\tau)
[
\widetilde\rho^I_{S,\tau}[\Sigma]
\hat A^I_\beta(y),
\hat A^I_\alpha(x)
]
\Big)
\nonumber\\
&+
\mathcal O(g^3),
\label{eq:mainTCL}
\end{align}
where the dressed relational memory kernel takes the factorized form
\begin{equation}
\mathcal C^{(\sigma)}_{\alpha\beta}(x,y;\tau)
=
\chi_\sigma(x,y;\tau)
G_{\alpha\beta}(x,y),
\label{eq:kernel}
\end{equation}
provided the clock and environmental sectors are statistically independent. Here
\begin{equation}
G_{\alpha\beta}(x,y)
=
\Tr_B
\!\left(
\rho_B
\hat B^I_\alpha(x)
\hat B^I_\beta(y)
\right)
\end{equation}
is the environmental correlation function and
$\chi_\sigma(x,y;\tau)$ denotes the clock correlation function associated with the finite temporal resolution of the relational clock. As shown in the Supplemental Material, Eq.~(\ref{eq:app-bathkernel}) follows from the statistical independence of clock fluctuations and environmental noise. More generally, when clock and environmental degrees of freedom are correlated, the reduced dynamics are governed directly by the corresponding composite clock-environment correlator.

Equation~\eqref{eq:kernel} makes explicit that the reduced dynamics possesses two distinct microscopic sources of memory. The factor $G_{\alpha\beta}(x,y)$ encodes the usual environmental spacetime correlations, while the clock correlator $\chi_\sigma(x,y;\tau)$ introduces an additional operational temporal smearing associated with finite-resolution relational clocks. Relativistic non-Markovianity therefore emerges jointly from environmental correlations and fluctuations in the operational notion of relational time itself.

The microscopic derivation of Eq.~\eqref{eq:kernel} from a composite clock--environment correlator, together with the conditions under which the factorized structure holds, is provided in the Supplemental Material.

The causal integration domain $J^-(x)\cap\Omega(\Sigma_0,\Sigma)$ ensures that the relational memory kernel depends only on the causal past of the spacetime point $x$, so that the reduced dynamics remains compatible with relativistic causal structure at the microscopic level.

Equation~\eqref{eq:mainTCL} constitutes the central result of the present work. Although local in the microscopic interaction densities, the reduced dynamics becomes temporally nonlocal in relational time through the causal memory kernel. The resulting non-Markovianity emerges jointly from relativistic environmental correlations and finite clock resolution, and therefore persists even when the underlying microscopic subsystem--environment interaction is local.

\paragraph{Emergent Markovian limit.—}
The non-Markovianity encoded in Eq.~\eqref{eq:mainTCL} has two distinct origins: relativistic environmental correlations and finite clock resolution. Because conditioning is performed with respect to a finite-width clock response function, relational time is operationally smeared over neighboring clock readings. The reduced dynamics therefore retains memory of both prior subsystem--environment interactions and relational clock fluctuations.

To recover an effective Markovian description, we introduce a smooth timelike scalar field $T(x)$ representing the dominant relational clock profile. Its normalized gradient,
\begin{equation}
u^\mu(x)
=
-
\frac{
\nabla^\mu T(x)
}{
\sqrt{
-\nabla^\nu T(x)\nabla_\nu T(x)
}
},
\label{eq:clockflow}
\end{equation}
defines a local timelike clock flow. The parameter $s$ denotes relational time measured along the integral curves of $u^\mu(x)$.

The existence of the timelike flow $u^\mu(x)$ provides the operational criterion by which the clock field becomes identifiable as a physical notion of time. Only relational observables whose correlations define an approximately stable timelike congruence admit an effective temporal interpretation. In this sense, temporal ordering emerges dynamically from coherent relational clock correlations rather than from an underlying external time parameter.

We assume a separation of scales,
\begin{equation}
\sigma
\gg
\tau_B,
\label{eq:separation}
\end{equation}
where $\tau_B$ is the bath correlation time and $\sigma$ is the clock resolution scale. Under approximate stationarity along the clock flow together with a local secular approximation, the interaction-picture operators admit a local frequency decomposition,
\begin{equation}
\hat A^I_\alpha(x;s)
=
\sum_\omega
e^{-i\omega s}
\hat A_\alpha(\omega;x),
\label{eq:localfreq}
\end{equation}
where $\hat A_\alpha(\omega;x)$ are local frequency components defined relative to the relational clock flow.

Because the dressed kernel decays rapidly compared to the relational coarse-graining scale, the causal memory integral may be extended effectively to infinite relational time. Substituting Eq.~\eqref{eq:localfreq} into Eq.~\eqref{eq:mainTCL} and retaining only resonant contributions yields
\begin{align}
\frac{\delta \overline\rho_{S,\tau}[\Sigma]}
{\delta\Sigma(x)}
={}&
-i
[
\hat h_S(x)
+
\hat h_{\rm LS}^{(\sigma)}(x),
\overline\rho_{S,\tau}[\Sigma]
]
\nonumber\\
&+
\sum_{\omega,\alpha,\beta}
\gamma^{(\sigma)}_{\alpha\beta}(\omega;x)
\Big(
\hat A_\beta(\omega;x)
\overline\rho_{S,\tau}[\Sigma]
\hat A_\alpha^\dagger(\omega;x)
\nonumber\\
&-
\frac12
\{
\hat A_\alpha^\dagger(\omega;x)
\hat A_\beta(\omega;x),
\overline\rho_{S,\tau}[\Sigma]
\}
\Big),
\label{eq:GKLS}
\end{align}
where $\overline\rho_{S,\tau}[\Sigma]$ denotes the coarse-grained conditioned state and $\hat h_{\rm LS}^{(\sigma)}(x)$ is the relational Lamb-shift Hamiltonian.

The dissipative coefficients are
\begin{equation}
\gamma^{(\sigma)}_{\alpha\beta}(\omega;x)
=
g^2
\int_{-\infty}^{\infty}
\dd s\,
e^{i\omega s}
\chi_\sigma(s;x)
G_{\alpha\beta}(s;x).
\label{eq:gamma}
\end{equation}
If the dressed kernel is of positive type, the matrix
$
\gamma^{(\sigma)}_{\alpha\beta}(\omega;x)
$
is positive semidefinite and Eq.~\eqref{eq:GKLS} takes GKLS form.

Equation~\eqref{eq:GKLS} therefore does not represent the fundamental relativistic dynamics. Rather, it emerges only after relational coarse-graining together with a separation of scales between bath memory and clock resolution. Markovianity consequently appears as an effective approximation rather than a microscopic principle of relativistic open-system dynamics.

\paragraph{Recovery of the AH limit.—}
The relational master equation~\eqref{eq:mainTCL} contains the Anastopoulos--Hu (AH) gravitational decoherence equation~\cite{AnastopoulosHu2013} as a limiting case corresponding to the recovery of an effectively classical notion of time.

Restricting to Minkowski spacetime with constant-time hypersurfaces,
$
t=\mathrm{const},
$
reduces the Schwinger--Tomonaga evolution to ordinary Hamiltonian evolution with respect to the coordinate-time parameter $t$. We further consider the sharp-clock limit,
\begin{equation}
\chi_\sigma(x,y;\tau)
\rightarrow
\chi_{\rm sharp}(t_x-t_y),
\label{eq:sharpclock}
\end{equation}
where $\chi_{\rm sharp}(t_x-t_y)$ is strongly localized around $t_x=t_y$. In this regime, the temporal smearing associated with finite clock resolution becomes negligible, so that relational time reduces effectively to the external coordinate time parameter.

Specializing the interaction Hamiltonian density to the transverse--traceless matter--graviton coupling used in the AH construction identifies the system operators with the transverse--traceless matter stress tensor $\hat\tau_{ij}(x)$ and the environmental operators with transverse--traceless graviton modes. Equation~\eqref{eq:mainTCL} then reduces to
\begin{align}
\frac{\partial \hat\rho_t}{\partial t}
={}&
-i[\hat H,\hat\rho_t]
\nonumber\\
&-
\frac{\kappa}{4}
\int d^3x
\int d^3y
\int_0^t ds\,
\mathcal N_{ijkl}(x-y,s)
\nonumber\\
&\times
[
\hat\tau_{ij}(x),
[
\hat\tau_{kl}(y,-s),
\hat\rho_t
]
]
\nonumber\\
&+
\frac{i\kappa}{4}
\int d^3x
\int d^3y
\int_0^t ds\,
\mathcal D_{ijkl}(x-y,s)
\nonumber\\
&\times
[
\hat\tau_{ij}(x),
\{
\hat\tau_{kl}(y,-s),
\hat\rho_t
\}
],
\label{eq:AHexplicit}
\end{align}
where $\kappa=16\pi G$, $\mathcal N_{ijkl}$ and $\mathcal D_{ijkl}$ are respectively the graviton noise and dissipation kernels, and $\hat\tau_{kl}(y,-s)$ denotes backward interaction-picture evolution under the free matter Hamiltonian. The explicit derivation of Eq.~\eqref{eq:AHexplicit} is provided in the Supplemental Material.

The AH equation therefore corresponds to the limit in which relational clock fluctuations become negligible and time effectively behaves as an external classical parameter. Conventional coordinate-time master equations thus emerge as limiting cases of the more general covariant relational dynamics developed here.

\paragraph{Conclusion.—}
We developed a covariant relational framework for relativistic open quantum dynamics in which evolution is conditioned on finite-resolution physical clocks directly on spacetime hypersurfaces. Unlike conventional approaches based on external classical time parameters and exact GKLS semigroup evolution, the present framework treats time operationally as a physical relational degree of freedom.

Our main result is that finite clock resolution generically produces intrinsically non-Markovian reduced dynamics even for local microscopic interactions. Environmental correlations and relational clock fluctuations jointly generate the memory kernel, demonstrating that temporal nonlocality is not merely a consequence of environmental complexity, but originates from the operational structure of relativistic time itself.

The framework therefore identifies a microscopic mechanism for the breakdown of exact relativistic Markovianity. The familiar GKLS structure emerges only after relational coarse-graining and suppression of clock fluctuations, establishing Markovian semigroup evolution as an effective emergent limit rather than a fundamental principle of relativistic open-system dynamics.

More broadly, our results suggest that exact Markovianity may be incompatible with a fundamentally relational notion of time. Conventional GKLS dynamics assumes evolution with respect to an ideal external time parameter, whereas relativistic physics admits only operational notions of time defined through physical clocks. Once finite clock resolution is taken into account, temporal correlations contribute directly to the reduced dynamics through the relational clock kernel, providing a microscopic origin for memory effects beyond environmental correlations alone. In this sense, the present framework identifies a mechanism by which exact Markovian evolution emerges only as an effective ideal-clock limit rather than as a fundamental property of relativistic open quantum systems.

By recovering the Anastopoulos--Hu gravitational decoherence equation in the sharp-clock fixed-foliation limit, we further showed that conventional coordinate-time master equations arise when relational clock fluctuations become negligible and time behaves effectively classically.

These results suggest that intrinsic memory effects may be unavoidable whenever temporal ordering is operationally defined through physical clocks rather than imposed externally. Beyond relativistic open quantum systems, the framework may therefore provide a useful setting for studying decoherence, information flow, and quantum correlations in regimes where no exact classical notion of time is fundamentally available, including relativistic detector--field systems and potentially quantum-gravitational settings such as black-hole spacetimes.

\begin{acknowledgments}
One of the authors, P.~N., acknowledges support from the RPF Postdoctoral Fellowship Program.
\end{acknowledgments}

\bibliographystyle{apsrev4-2}
\bibliography{gw_phases}

\clearpage
\onecolumngrid
\section*{Supplemental Material}

This supplemental material derives the relational memory kernel appearing in Eq.~\eqref{eq:kernel} of the main text from a microscopic system—a clock-bath model—and shows explicitly how the Anastopoulos-Hu gravitational decoherence equation emerges in the sharp-clock limit.

\section{Microscopic origin of the relational memory kernel}

In the main text, the reduced relational dynamics were shown to be governed by the dressed kernel
\begin{equation}
\mathcal C^{(\sigma)}_{\alpha\beta}(x,y;\tau)
=
\chi_\sigma(x,y;\tau)\,
G_{\alpha\beta}(x,y),
\label{eq:app-Ceff}
\end{equation}
which incorporates both environmental correlations and finite clock resolution.

The derivation presented below shows that the relational memory kernel is not introduced phenomenologically, but arises directly from correlation functions of a composite clock--bath sector. The factorized structure in Eq.~\eqref{eq:app-Ceff} emerges only when clock fluctuations and environmental noise are statistically independent.

We consider the total Hilbert space
\begin{equation}
\mathcal H_{\rm tot}
=
\mathcal H_S
\otimes
\mathcal H_\Xi,
\end{equation}
where
\begin{equation}
\mathcal H_\Xi
=
\mathcal H_C
\otimes
\mathcal H_B
\end{equation}
contains both the clock sector $C$ and the environmental bath $B$.

To implement relational time operationally, we introduce a Hermitian clock observable $\hat T(x)$ acting on the clock Hilbert space together with bath operators $\hat B_\alpha(x)$ acting on the bath sector. We then define the composite environmental operator
\begin{equation}
\hat{\mathcal E}_\alpha(x;\tau)
=
F_\sigma(\hat T(x)-\tau(x))
\,\hat B_\alpha(x),
\label{eq:app-E}
\end{equation}
where $F_\sigma$ is a finite-resolution response function of width $\sigma$.

The interaction Hamiltonian density is
\begin{equation}
\hat h_I(x;\tau)
=
g
\sum_\alpha
\hat A_\alpha(x)
\otimes
\hat{\mathcal E}_\alpha(x;\tau),
\label{eq:app-hi}
\end{equation}
where $\hat A_\alpha(x)$ denotes system operators and $g$ is the coupling constant.

The factor
\begin{equation}
F_\sigma(\hat T(x)-\tau(x))
\end{equation}
acts as a relational gate controlled by the clock degree of freedom. The interaction with the bath operator $\hat B_\alpha(x)$ therefore becomes effective only when the physical clock approximately registers the relational time $\tau(x)$ within the temporal resolution scale $\sigma$.

For concreteness, one may choose a Gaussian response profile
\begin{equation}
F_\sigma(u)
=
\frac{1}{\sqrt{2\pi}\sigma}
\exp\!\left(
-\frac{u^2}{2\sigma^2}
\right).
\label{eq:app-gaussian}
\end{equation}

We assume an initially factorized state
\begin{equation}
\rho_{\rm tot}[\Sigma_0]
=
\rho_S[\Sigma_0]
\otimes
\rho_\Xi,
\qquad
\rho_\Xi
=
\rho_C
\otimes
\rho_B,
\label{eq:app-rho0}
\end{equation}
where $\rho_C$ and $\rho_B$ denote respectively the initial clock and bath states.

Passing to the interaction picture with respect to the free system, bath, and clock evolution yields
\begin{equation}
\frac{\delta \rho^I_{\rm tot}[\Sigma]}
{\delta\Sigma(x)}
=
-i
\big[
\hat h_I^I(x;\tau),
\rho^I_{\rm tot}[\Sigma]
\big],
\label{eq:app-STI}
\end{equation}
where
\begin{equation}
\hat h_I^I(x;\tau)
=
g
\sum_\alpha
\hat A_\alpha^I(x)
\otimes
\hat{\mathcal E}_\alpha^I(x;\tau).
\label{eq:app-hII}
\end{equation}

Iterating Eq.~\eqref{eq:app-STI} generates the Dyson expansion
\begin{align}
\rho^I_{\rm tot}[\Sigma]
={}&
\rho_0^I
-
i
\int_{\Omega(\Sigma_0,\Sigma)}
d{\rm vol}_x\,
\big[
\hat h_I^I(x;\tau),
\rho_0^I
\big]
\nonumber\\
&-
\int_{\Omega(\Sigma_0,\Sigma)}
d{\rm vol}_x
\int_{\Omega(\Sigma_0,x)}
d{\rm vol}_y
\nonumber\\
&\times
\bigg[
\hat h_I^I(x;\tau),
\big[
\hat h_I^I(y;\tau),
\rho_0^I
\big]
\bigg]
\nonumber\\
&+
\mathcal O(g^3),
\label{eq:app-dyson}
\end{align}
where $\Omega(\Sigma_0,\Sigma)$ denotes the spacetime region bounded by the hypersurfaces $\Sigma_0$ and $\Sigma$.

The reduced relational state is obtained by tracing over the clock and bath sectors,
\begin{equation}
\rho_S^I[\Sigma;\tau]
=
\Tr_\Xi
\Big(
\rho^I_{\rm tot}[\Sigma]
\Big).
\end{equation}

Assuming vanishing one-point functions,
\begin{equation}
\langle
\hat{\mathcal E}_\alpha^I(x;\tau)
\rangle_\Xi
=
0,
\end{equation}
the first-order contribution vanishes. The second-order contribution becomes
\begin{align}
\rho_{S,(2)}^I[\Sigma;\tau]
={}&
-
g^2
\sum_{\alpha,\beta}
\int
d{\rm vol}_x
\int
d{\rm vol}_y
\,
\Big(
\nonumber\\
&
W_{\alpha\beta}^{>}(x,y;\tau)
\big[
\hat A_\alpha^I(x),
\hat A_\beta^I(y)\rho_S^I
\big]
\nonumber\\
&+
W_{\beta\alpha}^{<}(y,x;\tau)
\big[
\rho_S^I\hat A_\beta^I(y),
\hat A_\alpha^I(x)
\big]
\Big),
\label{eq:app-second}
\end{align}
where
\begin{align}
W_{\alpha\beta}^{>}(x,y;\tau)
&=
\Big\langle
\hat{\mathcal E}_\alpha^I(x;\tau)
\hat{\mathcal E}_\beta^I(y;\tau)
\Big\rangle_\Xi,
\\
W_{\alpha\beta}^{<}(x,y;\tau)
&=
\Big\langle
\hat{\mathcal E}_\beta^I(y;\tau)
\hat{\mathcal E}_\alpha^I(x;\tau)
\Big\rangle_\Xi.
\end{align}

Using Eq.~\eqref{eq:app-E}, the composite correlator becomes
\begin{align}
W_{\alpha\beta}^{>}(x,y;\tau)
={}&
\Big\langle
F_\sigma(\hat T(x)-\tau(x))
\hat B_\alpha(x)
\nonumber\\
&
\times
F_\sigma(\hat T(y)-\tau(y))
\hat B_\beta(y)
\Big\rangle_\Xi.
\label{eq:app-expand}
\end{align}

If the clock and bath sectors are statistically independent,
\begin{equation}
\rho_\Xi
=
\rho_C
\otimes
\rho_B,
\end{equation}
then the expectation value factorizes:
\begin{align}
W_{\alpha\beta}^{>}(x,y;\tau)
={}&
\Big\langle
F_\sigma(\hat T(x)-\tau(x))
F_\sigma(\hat T(y)-\tau(y))
\Big\rangle_C
\nonumber\\
&
\times
\Big\langle
\hat B_\alpha(x)
\hat B_\beta(y)
\Big\rangle_B.
\label{eq:app-factorstep}
\end{align}

We therefore define the clock correlation function
\begin{equation}
\chi_\sigma(x,y;\tau)
:=
\Big\langle
F_\sigma(\hat T(x)-\tau(x))
F_\sigma(\hat T(y)-\tau(y))
\Big\rangle_C,
\label{eq:app-clockkernel}
\end{equation}
and the bath correlation function
\begin{equation}
G_{\alpha\beta}(x,y)
:=
\Big\langle
\hat B_\alpha(x)
\hat B_\beta(y)
\Big\rangle_B.
\label{eq:app-bathkernel}
\end{equation}

The composite correlator therefore factorizes as
\begin{equation}
W_{\alpha\beta}^{>}(x,y;\tau)
=
\chi_\sigma(x,y;\tau)
\,G_{\alpha\beta}(x,y).
\label{eq:app-factorized}
\end{equation}

An analogous relation holds for the lesser correlator,
\begin{equation}
W_{\alpha\beta}^{<}(x,y;\tau)
=
\chi_\sigma(x,y;\tau)
\,G_{\alpha\beta}^{<}(x,y).
\end{equation}

Equation~\eqref{eq:app-factorized} reproduces the dressed relational kernel introduced in the main text.

The important point is that the clock response function and the environmental memory kernel originate from a single microscopic composite correlator. The factorized structure therefore emerges only when clock fluctuations and bath noise are statistically independent.

More generally, if the same microscopic degrees of freedom contribute simultaneously to relational clock fluctuations and environmental noise, then the factorization
\begin{equation}
W_{\alpha\beta}
=
\chi_\sigma G_{\alpha\beta}
\end{equation}
breaks down. In that case the reduced dynamics is governed directly by the irreducible composite spacetime correlator itself.

Promoting the perturbative expansion to a local-in-state time-convolutionless form yields
\begin{align}
\frac{\delta \widetilde\rho^I_{S,\tau}[\Sigma]}
{\delta\Sigma(x)}
={}&
-g^2
\sum_{\alpha\beta}
\int_{J^-(x)\cap\Omega(\Sigma_0,\Sigma)}
\dd\mathrm{vol}_y
\Big(
\mathcal C^{(\sigma)}_{\alpha\beta}(x,y;\tau)
\nonumber\\
&\times
[
\hat A^I_\alpha(x),
\hat A^I_\beta(y)
\widetilde\rho^I_{S,\tau}[\Sigma]
]
\nonumber\\
&+
\mathcal C^{(\sigma)*}_{\beta\alpha}(y,x;\tau)
[
\widetilde\rho^I_{S,\tau}[\Sigma]
\hat A^I_\beta(y),
\hat A^I_\alpha(x)
]
\Big),
\label{eq:app-TCL}
\end{align}
where
\begin{equation}
\mathcal C^{(\sigma)}_{\alpha\beta}(x,y;\tau)
=
\chi_\sigma(x,y;\tau)
G_{\alpha\beta}(x,y).
\end{equation}

The integration region
\begin{equation}
J^-(x)\cap\Omega(\Sigma_0,\Sigma)
\end{equation}
ensures causal ordering of the spacetime evolution.

Equation~\eqref{eq:app-TCL} is generically non-Markovian because the conditioned state at the spacetime point $x$ depends on the causal history encoded in the dressed relational kernel.

\section{Emergent GKLS limit}

To recover an effective Markovian regime, we introduce a smooth timelike scalar field $T(x)$ representing the dominant relational clock profile. Throughout this work we adopt the mostly-plus metric convention $(-,+,+,+).$

The hypersurfaces
\begin{equation}
T(x)=\mathrm{const}
\end{equation}
therefore define a foliation of spacetime by spacelike slices. Since the gradient $\nabla_\mu T(x)$ is normal to these hypersurfaces, it is timelike and satisfies
\begin{equation}
\nabla^\mu T\nabla_\mu T<0.
\end{equation}

The associated future-directed unit clock-flow vector field is obtained by normalizing the gradient,
\begin{equation}
u^\mu(x)
=
-
\frac{
\nabla^\mu T(x)
}{
\sqrt{
-\nabla^\nu T(x)\nabla_\nu T(x)
}
}.
\end{equation}
The minus sign ensures future-directed flow in the mostly-plus convention. The parameter $s$ denotes relational time measured along the integral curves of $u^\mu(x)$.

We now assume a separation of scales,
\begin{equation}
\sigma
\gg
\tau_B,
\end{equation}
where $\sigma$ is the clock resolution scale and $\tau_B$ is the characteristic bath correlation time. In this regime, short-time environmental correlations become unresolved relative to the relational coarse-graining scale defined by the clock.

Assuming approximate stationarity along the clock flow together with a local secular approximation, the interaction-picture operators admit the local frequency decomposition
\begin{equation}
\hat A^I_\alpha(x;s)
=
\sum_\omega
e^{-i\omega s}
\hat A_\alpha(\omega;x),
\end{equation}
where $\hat A_\alpha(\omega;x)$ denotes the local frequency component associated with the relational frequency $\omega$.

Under these assumptions, the dressed relational kernel decays sufficiently rapidly along the clock flow so that the causal memory integral may be extended effectively to infinite relational time. The dissipative coefficients then become
\begin{equation}
\gamma^{(\sigma)}_{\alpha\beta}(\omega;x)
=
g^2
\int_{-\infty}^{\infty}
\dd s\,
e^{i\omega s}
\chi_\sigma(s;x)
G_{\alpha\beta}(s;x),
\end{equation}
where $\chi_\sigma(s;x)$ is the coarse-grained clock kernel and $G_{\alpha\beta}(s;x)$ is the bath correlation function evaluated along the relational clock flow.

If the dressed kernel is of positive type, the matrix
$
\gamma^{(\sigma)}_{\alpha\beta}(\omega;x)
$
is positive semidefinite by Bochner's theorem and the reduced dynamics takes GKLS form.

The Markovian regime therefore emerges only after relational coarse-graining together with a separation of scales between bath memory and clock resolution. In this sense, GKLS dynamics appears as an effective long-time description of the more general relationally non-Markovian covariant dynamics developed in the main text.

\section{Recovery of the Anastopoulos--Hu equation}

We now specialize the general relational framework to linearized gravity in order to recover the Anastopoulos--Hu (AH) gravitational decoherence equation.

To recover the AH equation, we specialize the interaction Hamiltonian density to linearized gravity in the transverse-traceless gauge,
\begin{equation}
\hat h_I(x)
=
-\frac{\kappa}{2}
\hat\tau_{ij}(x)
\hat h^{ij}_{\rm TT}(x),
\label{eq:Sgrav}
\end{equation}
where $\kappa=16\pi G$, with $G$ the Newton gravitational constant, $\hat\tau_{ij}(x)$ denotes the transverse-traceless projection of the matter stress-tensor operator, and $\hat h^{ij}_{\rm TT}(x)$ is the transverse-traceless graviton field operator.

Restricting to Minkowski spacetime with constant-time hypersurfaces, $t=\mathrm{const},$ reduces the Schwinger-Tomonaga evolution to ordinary Hamiltonian evolution with respect to the coordinate-time parameter $t$.

Taking the sharp-clock limit,
\begin{equation}
\chi_\sigma(x,y;\tau)
\rightarrow
\chi_{\rm sharp}(t_x-t_y),
\end{equation}
suppresses relational clock fluctuations and restores an effectively classical notion of time.

The reduced master equation then becomes
\begin{align}
\frac{\partial \hat\rho_t}{\partial t}
={}&
-i[\hat H,\hat\rho_t]
\nonumber\\
&-
\frac{\kappa}{4}
\int_0^t ds
\int d^3x
\int d^3y
\nonumber\\
&\times
\Big(
G_{ijkl}(x-y,s)
[
\hat\tau_{ij}(x),
\hat\tau_{kl}(y,-s)\hat\rho_t
]
\nonumber\\
&+
G^*_{ijkl}(x-y,s)
[
\hat\rho_t\hat\tau_{kl}(y,-s),
\hat\tau_{ij}(x)
]
\Big),
\label{eq:AHpre}
\end{align}
where
\begin{equation}
G_{ijkl}(x-y,s)
=
\mathrm{Tr}_B
\left(
\rho_B
\hat h^{ij}_{\rm TT}(x,t)
\hat h^{kl}_{\rm TT}(y,t-s)
\right)
\end{equation}
is the graviton two-point correlation function.

The interaction-picture operator
\begin{equation}
\hat\tau_{kl}(y,-s)
=
e^{-i\hat H s}
\hat\tau_{kl}(y)
e^{i\hat H s}
\end{equation}
denotes backward free evolution under the matter Hamiltonian $\hat H$.

The graviton correlator may be decomposed into symmetric and antisymmetric parts,
\begin{equation}
G_{ijkl}(x-y,s)
=
\mathcal N_{ijkl}(x-y,s)
+
\frac{i}{2}
\mathcal D_{ijkl}(x-y,s),
\label{eq:Sdecomp}
\end{equation}
where
\begin{align}
\mathcal N_{ijkl}(x-y,s)
={}&
\frac12
\left(
G_{ijkl}(x-y,s)
+
G_{klij}(y-x,-s)
\right),
\\
\mathcal D_{ijkl}(x-y,s)
={}&
-i
\left(
G_{ijkl}(x-y,s)
-
G_{klij}(y-x,-s)
\right).
\end{align}

The kernel $\mathcal N_{ijkl}(x-y,s)$ is the symmetrized graviton correlation function responsible for gravitational decoherence, while $\mathcal D_{ijkl}(x-y,s)$ is the causal dissipation kernel associated with gravitational backreaction.

Substituting Eq.~\eqref{eq:Sdecomp} into Eq.~\eqref{eq:AHpre} yields
\begin{align}
\frac{\partial \hat\rho_t}{\partial t}
={}&
-i[\hat H,\hat\rho_t]
\nonumber\\
&-
\frac{\kappa}{4}
\int d^3x
\int d^3y
\int_0^t ds\,
\mathcal N_{ijkl}(x-y,s)
\nonumber\\
&\times
[
\hat\tau_{ij}(x),
[
\hat\tau_{kl}(y,-s),
\hat\rho_t
]
]
\nonumber\\
&+
\frac{i\kappa}{4}
\int d^3x
\int d^3y
\int_0^t ds\,
\mathcal D_{ijkl}(x-y,s)
\nonumber\\
&\times
[
\hat\tau_{ij}(x),
\{
\hat\tau_{kl}(y,-s),
\hat\rho_t
\}
].
\label{eq:AHfinal}
\end{align}

Equation~\eqref{eq:AHfinal} reproduces the Anastopoulos--Hu gravitational decoherence equation. The first nonunitary contribution proportional to the noise kernel generates gravitational decoherence, while the second contribution involving the dissipation kernel describes the corresponding dissipative backreaction.

The AH limit therefore corresponds to the regime in which relational temporal fluctuations become negligible and the operational notion of time reduces effectively to an external classical parameter. Equation~\eqref{eq:AHfinal} thus emerges as the sharp-clock limit of the more general covariant relational dynamics developed in the main text.

\end{document}